\documentclass[iop, revtex4-2]{emulateapj}

\usepackage{color}
\usepackage{courier}
\usepackage[caption=false]{subfig}
\usepackage{hyperref}

\shortauthors{Smith, Tandon \& Wagoner}
\shorttitle{High-Frequency QPO Models}

\begin{document}

\title{Confrontation of Observation and Theory: High Frequency QPOs in X-ray Binaries, Tidal Disruption Events, and Active Galactic Nuclei}

\author{Krista Lynne Smith\altaffilmark{1,2,3}, Celia R. Tandon \altaffilmark{4}, Robert V. Wagoner\altaffilmark{4}}

\altaffiltext{1}{Department of Physics, Southern Methodist University, kristas@smu.edu}
\altaffiltext{2}{KIPAC at SLAC, Stanford University}
\altaffiltext{3}{Einstein Fellow}
\altaffiltext{4}{Department of Physics and KIPAC, Stanford University}

\begin{abstract}
We compile observations of high-frequency quasi-periodic oscillations (QPOs) around black holes, both stellar and supermassive, and compare their positions in the parameter space of black hole mass, spin, and oscillation frequency. We find that supermassive black holes occupy a separate region of parameter space than stellar, and further, that QPOs seen around tidal disruption events rather than Seyfert-type AGN occupy an entirely different space. We then compare these results to the orbital resonance, diskoseismic, warped disk, and disk-jet coupling theoretical models for the origin of high-frequency QPOs. We find that while oscillations around stellar mass black holes are generally consistent with the above models, supermassive black holes are decidedly not. Oscillations seen in tidal disruption events are consistent with oscillations near the frequency of the innermost stable circular orbit (ISCO), while QPOs in AGN are not accounted for by any of the physical models in consideration. This indicates that despite the scale invariance of accretion processes implied by a decades-wide correlation between QPO frequency and black hole mass, any theory of high frequency QPOs must relate the frequency to more than just the mass and spin.

\end{abstract}


\section{Introduction}

Quasi-periodic oscillations (QPOs) in X-ray flux light curves have long been observed in X-ray binaries hosting stellar-mass black holes \citep[e.g., ][]{Remillard2006}. These variations are likely to arise from quite near the black hole itself, and exhibit frequencies that scale inversely with the black hole mass. For high-frequency QPOs (HFQPOs), which occur on timescales of a few hundred Hz, the oscillation frequencies are usually stable and often in a 3:2 ratio. The phenomenon is not universal, however. Rather, HFQPOs are elusive, with low duty cycle, and appearing in only 11 out of $\sim7000$ observations of 22 stellar mass black holes \citep{Belloni2012}. The oscillations occur only in certain states of luminosity and hardness; in X-ray binaries, HFQPOs occur only in the steep power law state \citep{Remillard2006} or ``anomalous" high-soft state \citep{Belloni2012,Motta2016}; both of these correspond to a luminous state with a soft X-ray spectrum including a thermal disk component. The amplitude of the modulation also appears to increase with increasing photon energy \citep{Remillard2006}.

The similar phenomenon of low-frequency QPOs occurs at a few to dozens of Hz, but have time-varying frequency centroids and are likely due to other phenomena. Quasi-periods have also been observed in supermassive black hole light curves, first by \citet{Gierlinski2008}, and more often in a recent proliferation of such discoveries as people re-examine archival data from the XMM-\emph{Newton} satellite taken for other purposes \citep[e.g.][]{Lin2013,Alston2015,Zhang2020}. Some QPOs have been observed in extreme-UV and optical AGN light curves as well \citep{Halpern2003,Smith2018b}. The QPOs in supermassive sources are not expected to vary over observational timescales, since the propagation time of viscous instabilities and accretion rate fluctuations is very long compared to stellar mass sources, from tens to hundreds of years. 

An impressive correlation between the frequency of QPOs (both high and low) and the black hole mass has been observed spanning many orders of magnitude, from stellar mass to supermassive black hole sources  \citep{Abramowicz2004,Zhou2015}. This is a strong argument for scale invariance of the accretion process, and suggests that the mechanism responsible for QPOs may be the same in both types of object. It also raises the tantalizing possibility of using observed periodicities to measure black hole masses; any new and independent method for weighing supermassive black holes would be quite valuable, particularly as we gather progressively larger and finer-sampled light curves of AGN with instruments like eROSITA, TESS, and future ground-based surveys.

Despite the exciting possibilities raised by the observations, there is still no consensus on the physical mechanism underlying QPOs, either high- or low-frequency. A number of candidate theories have been proposed, most of which concern resonant, epicyclic, or orbital frequencies dictated by the gravitational field of the black hole and thus explain the observed mass scaling. However, none have yet successfully reproduced all of the observed features of QPOs or the properties of individual objects. 

In this paper, we compare the high-frequency QPOs seen in stellar and supermassive black hole sources to several theoretical models for their origin. In Section~\ref{sec:theories}, we briefly review the theoretical models in consideration. In Section~\ref{sec:observations}, we show how observed QPOs from the literature compare to the models. In Section~\ref{sec:discussion} we discuss the physical interpretation and consequences of our findings, as well as possible selection effects. We conclude with a brief summary in Section~\ref{sec:conclusion}.

\section{Theories}
\label{sec:theories}
One might expect that the stable frequencies observed in HFQPOs should be governed most strongly by the fixed gravitational field of the black hole, rather than properties of the accretion disk or the X-ray emitting corona, which are known to fluctuate. The equatorial orbital, epicyclic, and black hole spin frequencies are inversely proportional to the mass of the black hole.
Therefore, one might expect that black hole HFQPO frequencies $f$ would obey a dimensionless relation of the form $(GM/c^3)f = F(a,X)$, where $a = cJ/(GM^2)$ is the dimensionless black hole spin parameter and $X$ includes dimensionless non-gravitational properties such as the accretion rate, often parameterized by the luminosity ratio $L/L_{Edd}$. Since the luminosity of the stellar mass HFQPO sources is time dependent, and the observed frequencies generally are not, the dependence of $f$ on this component of $X$ must be weak. We do note, however, that recently \citet{Belloni2019} found evidence for slight variability in the HFQPO frequency of GRS~1915+105 that appears to depend on the spectral hardness.

\begin{figure}[tbp]
\plotone{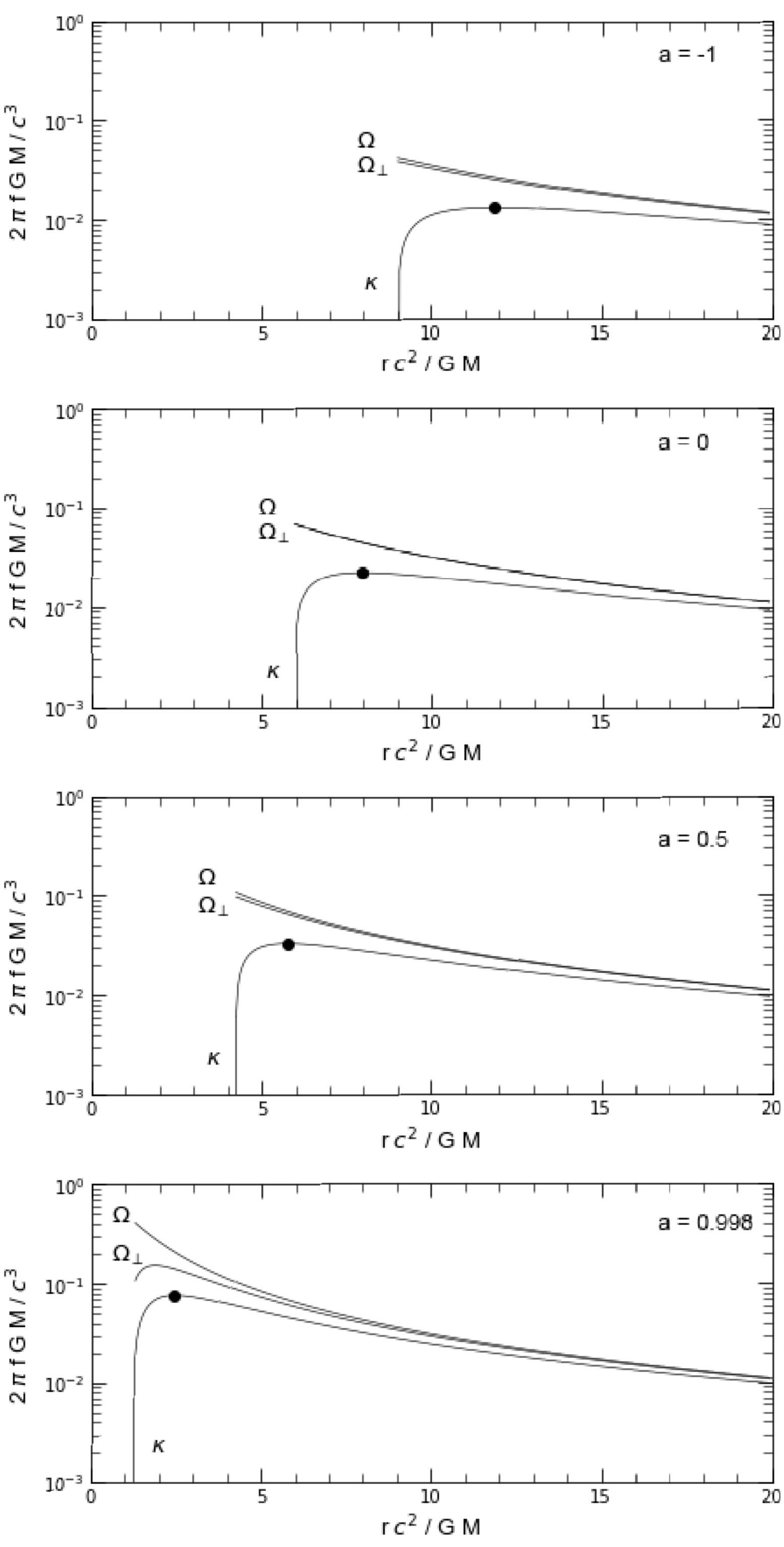}
\caption{The circular orbital ($\Omega$), vertical epicyclic ($\Omega_{\perp}$), and radial epicyclic ($\kappa$) frequencies of free particles in the equatorial plane of the black hole, for four values of the black hole dimensionless angular momentum. The maximum value of $\kappa$ is indicated by a black dot.}
\label{fig:orbitals}
\end{figure}

\citet{Kato1980} pointed out that in general relativity, in contrast to Newtonian gravity, the radial epicyclic frequency $\kappa$ does not simply increase with decreasing radial distance to a black hole, but instead reaches a maximum at a few gravitational radii and then decreases to zero at the innermost stable circular orbit (ISCO), see Figure~\ref{fig:orbitals}. Consequently, oscillations in this region of the disk may become ``trapped," and result in observable periodic flux variations. 

We shall consider the following classes of theoretical models for the origins of HFQPOs around black holes.

\subsection{Orbital Resonances}
\label{sec:orbres}
The original resonance models invoked nonlinear coupling between the orbital frequency and the radial epicyclic frequency \citep{Abramowicz2001}, or the between the vertical and radial epicyclic frequencies \citep{Abramowicz2003} in the accretion disk. In the latter case, a coupling was assumed when the frequencies were in a 3:2 ratio, in agreement with the peaks seen in a handful of X-ray power spectra known at that time.

To investigate a physical basis for this coupling, \citet{Horak2008} analyzed weak nonlinear interactions between the epicyclic modes in slender tori, extending earlier work of \citet{Abramowicz2003}. He found that the strongest resonance was between the axisymmetric modes when their frequencies were in a 3:2 ratio.


\subsection{Diskoseismology}
\label{sec:diskoseismology}

Astrophysical seismic modes are probably most familiar from studies of the sun and stellar interiors. Acoustic pressure ($p$) modes have been observed in the Sun and other stars. Gravity ($g$) modes are predicted to exist as well, driven by buoyancy. Accretion disks are subject to the same phenomena, although the excitation methods are different: turbulence is magnetically driven rather than convective, and the $g$-modes are governed by inertial forces rather than buoyancy. Theoretical expectations for these modes in accretion disks are  summarized by \citet{Wagoner2001}, who tabulate the comparative radial locations and extents of each of the fundamental modes, which bear on their observability in that larger modes with less leakage past the ISCO will have more radiative flux to modulate.

The most visible should be the fundamental $g$-mode, since it has a relatively large extent ($\Delta r \approx GM/c^2$) and is located at the temperature maximum of the disk, away from leakage through the ISCO. It is trapped just below the maximum of the radial epicyclic frequency (indicated by the black dots in Figure~\ref{fig:orbitals}). Its frequency $f$  obeys the relation 
           \[ Mf/10M_\sun = F(a)[1-0.1L/L_{Edd}] \]
for $L\lesssim 0.5 L_{Edd}$, with $F = 71 - 246$ Hz for $a=0-1$ \citep{Perez1997,Wagoner2001}. This model garners further plausibility from the fact that it, alone among the diskoseismic modes, has been unambiguously detected in a hydrodynamic disk simulation \citep{Reynolds2009}; however, we note that in this study, the QPO disappeared when a magnetic field was included, due to turbulence induced by the magneto-rotational instability. Very recent work by \citet{Dewberry2020} presents a more optimistic picture: when eccentricity or warping is forced in relativistic magnetohydrodynamic simulations, the coupling of the trapped $g$- and $p$-modes to the distortion amplifies the modes even in the presence of MHD turbulence. 

The pressure $p$-modes can be trapped between the radial epicyclic frequency and the ISCO, but have a much smaller radial extent in the disk. They may also be more prone to leakage and infall, further reducing their potential observability. However, in simulations with hydrodynamic viscosity, the $g$-mode leaked into outgoing $p$-waves of the same frequency, and propagated at the trapped-mode frequency out to larger disk radii \citep{Oneill2009}. Such an occurrence would lead to a larger observed luminosity modulation. Furthermore, some non-axisymmetric, low-order $p$-modes may be amplified by the co-rotation resonance between the outer and inner Lindblad resonances, or by certain magnetic field conditions in MHD disks \citep{Lai2013}; the lower limit of the frequencies of these modes is $\approx0.5\times f_\mathrm{ISCO}$. 

One drawback of diskoseismic models is the inability to account for the observed 3:2 ratio of HFQPO peaks seen in stellar mass black holes; no coupled modes in the proper ratio have been predicted.

\subsection{Warped Disks}
\label{sec:warp}
An interaction between a warped disk (with vertical displacement proportional to $\sin(\varphi)$) and a normal mode of oscillation can be excited at the radius $r_*$ where $\kappa = \Omega/2$ \citep{Kato2008}. This can produce oscillations at frequencies $\kappa(r_*)$, $2\kappa(r_*)$, and $3\kappa(r_*)$. Since $r_*$ is close to the radius of the maximum value of $\kappa$, the lowest frequency is almost equal to that of the $g$-mode referred to above, for all values of the black hole spin $a$; see figure 12.12 of \citet{Kato2008} for the dependence of $2\kappa(r_*)$ on $a$, but note the factor of two error in their Equation~12.25. 


\subsection{Disk-Jet Coupling}
\label{sec:diskjet}

A different important candidate for exciting non-axisymmetric modes in accretion flows around black holes is fluid instability between the magnetic field surrounding the accreting object and the inflowing material. \citet{Li2004} evaluated the interface between an accretion disk and the magnetosphere of the central object, and found that the boundary gives rise to both Rayleigh-Taylor and Kelvin-Helmholtz instabilities. In turn, these enable unstable modes to grow in the disk, resulting in non-axisymmetric patterns that may translate into quasi-periodic brightness variations. This model is also interesting in that it naturally describes the frequency width and finite lifetimes of QPOs, and may even explain some frequency ratios between QPO peaks, since the low wavenumber modes will dominate due to the radial dependence of the perturbation.

\citet{McKinney2012} expanded this model in MHD simulations, and found that a Raleigh-Taylor instability developed at the interface of the disk and jet regions in magnetically-choked accretion flows. Oscillations were indeed induced at the magnetospheric interface. At this location, the rotational frequencies of the plasma, $\Omega$, and the field lines, $\Omega_F$, are forced to be similar by the strength of the magnetic field. The QPO frequency is therefore set by the black hole spin, which drags the field lines with $\Omega_F \sim \Omega_\mathrm{BH} / 4$, where the spin frequency $\Omega_{\mathrm{BH}} = (c^3/GM) a / [ 2(1+\sqrt{1-a^2}) ] $. A robust detection of a HFQPO  was reported. This result is likely to be valid only for black hole spins $a\gtrsim 0.4$, below which the plasma disk rotation may dominate over the spin frequency.

\subsection{Tori, Hotspots, and Precession}
\label{sec:precession}
This class of theories produces relations of the form $Mf = F(a,r)$, where $r$ is the radial position of the pressure maximum of the torus \citep{Rezzolla2003} or the center of the orbiting `hot spot'. However, a hot spot overdensity would be subject to destruction  by the shear of the disk. The relativistic precession model \citep{Stella1999} involves the nodal precession frequency $\Omega - \Omega_\perp$ and the periastron precession frequency $\Omega - \kappa$, but no resonance is involved. 

\begin{figure*}
\begin{tabular}{cc}
  \includegraphics[width=0.5\textwidth]{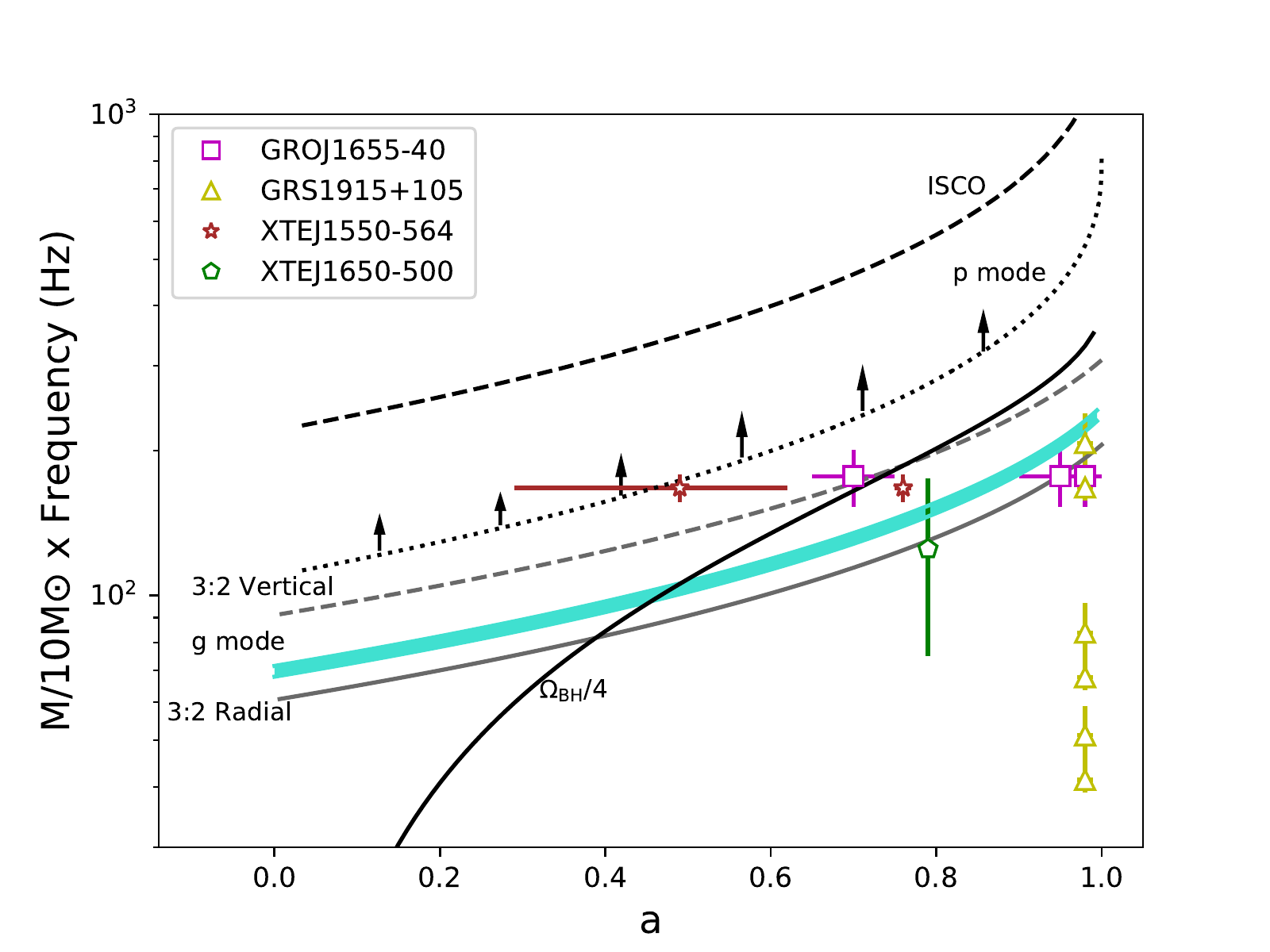}   & \includegraphics[width=0.51\textwidth]{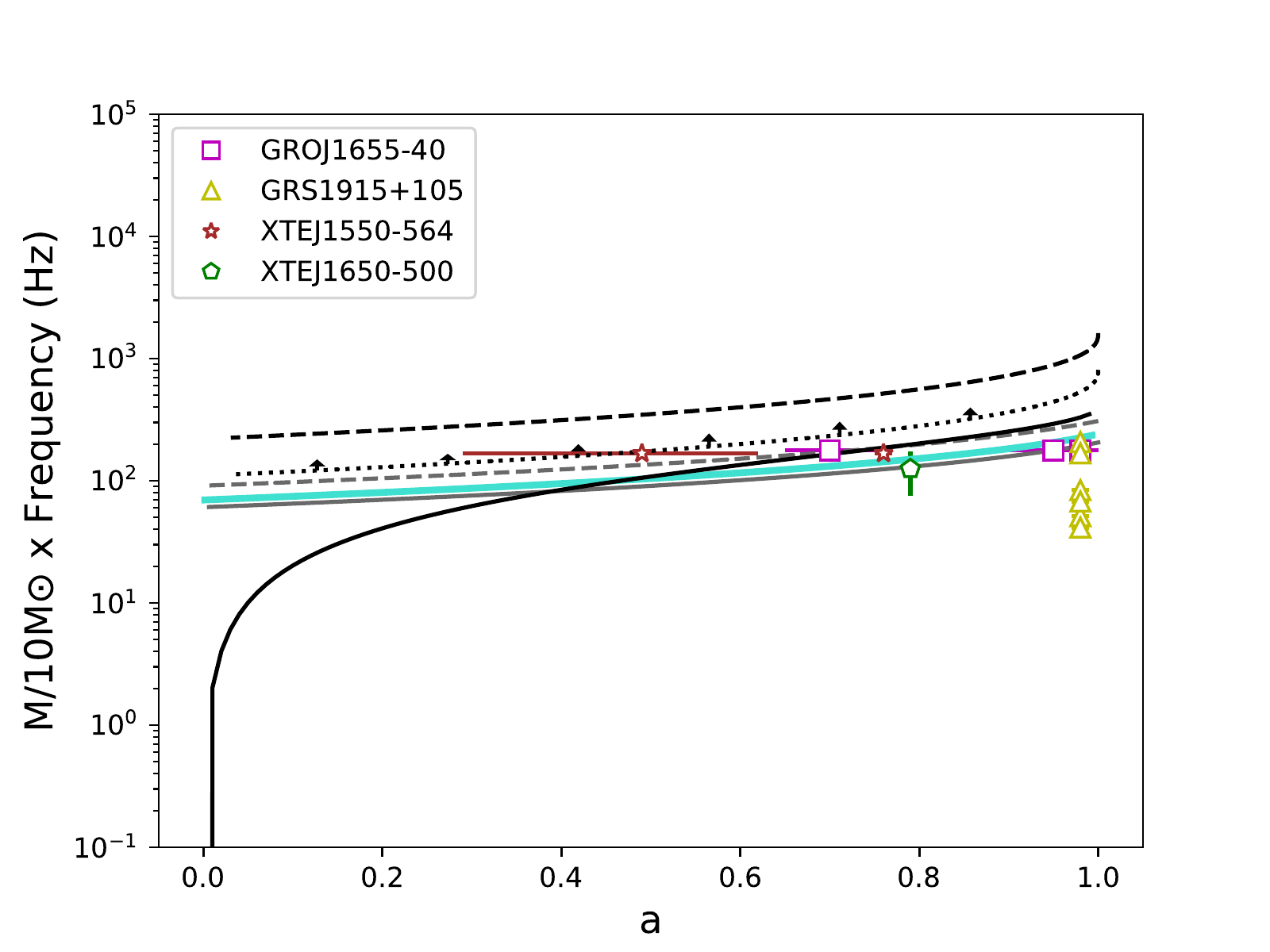} \\

\end{tabular}
\caption{Relationship between the product of mass and frequency, and the dimensionless spin parameter, for stellar mass black holes (left), and again with the same axis scaling as in Figure~\ref{fig:smbh_obs} (right). The curves show several theoretical models as labelled in the left panel, from top to bottom: the black dashed line indicates an orbital frequency at the ISCO; the dotted line indicates the lower limit for non-axisymmetric $p$-modes from \citet{Lai2013}; the grey dashed and solid lines indicate the 3:2 vertical and radial epicyclic frequencies, respectively; the turquoise shaded region indicates the $g$-mode (with its thickness corresponding to $L <  0.5L_{Edd}$); and the solid black line indicates QPOs formed by fluid instabilities at the disk/jet interface. For objects that exhibit a 3:2 ratio of QPO peaks, only the lower frequency is shown. Multiple values for individual objects indicate different epochs or black hole mass or spin measurement values. References for plotted values can be found in Table~\ref{t:xrb_obs}.}
\label{fig:xrb_obs}
\end{figure*}

\begin{figure*}
   \centering
    \includegraphics[width=\textwidth]{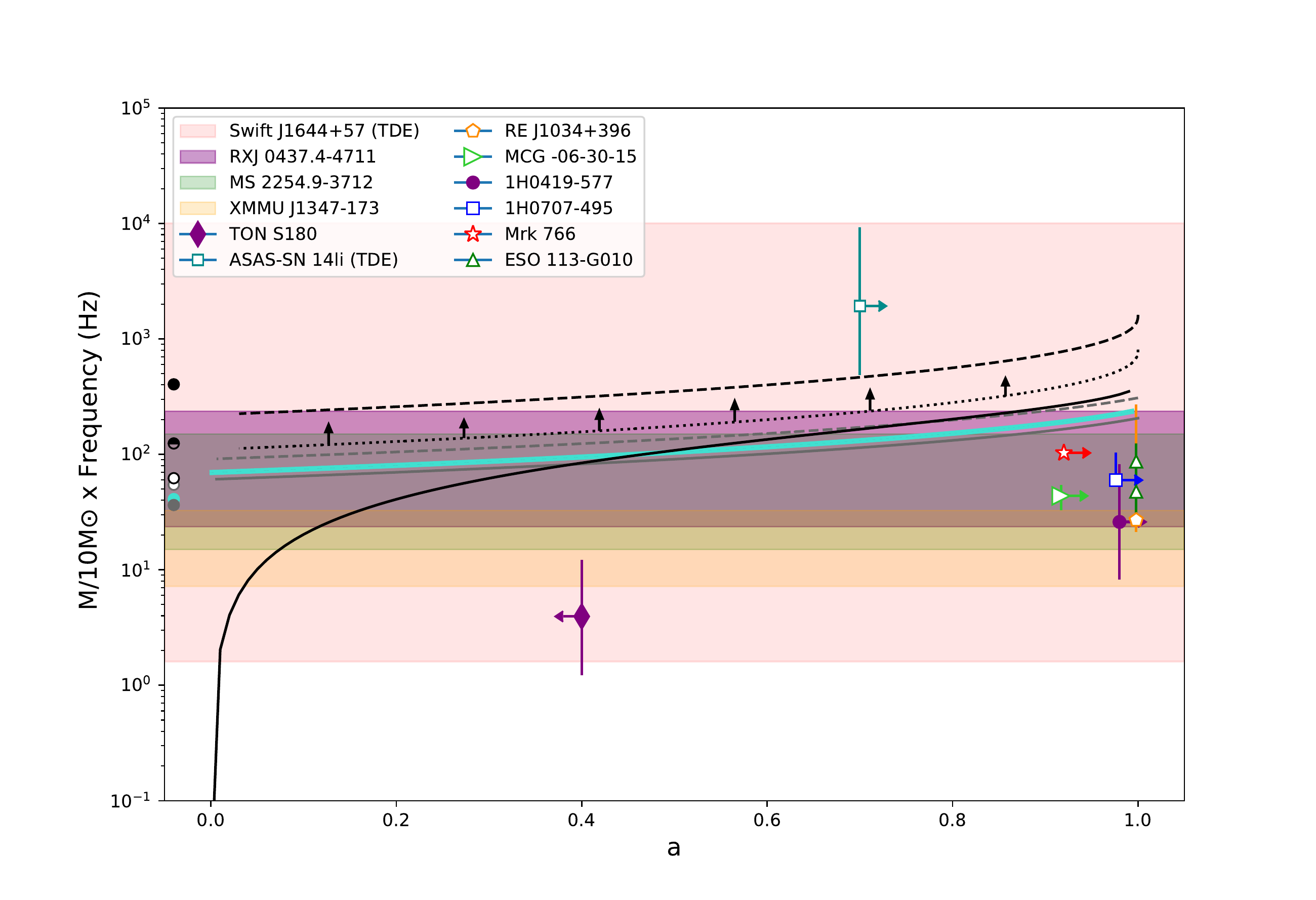}
    \caption{Relationship between the product of mass and frequency, and the dimensionless spin parameter, for supermassive black holes. Frequencies are in the galaxy rest frame, corrected for redshift. Data points indicate objects for which the spin is known; most of these have only spin lower limits, indicated by rightward arrows. Shaded regions indicate objects with mass estimates, often with large errors, but no spin estimates in the literature. Purple symbols and bands denote the only detections not found in X-rays, the EUV QPO candidates RXJ~0437.4-4711, TON~S180, and 1H~0419-577. Curves are the same as those labelled in Figure~\ref{fig:xrb_obs}, with their extrapolated values for $a=-1$ given on the left axis. References for the plotted values can be found in Table~\ref{t:smbh_obs}.}
    \label{fig:smbh_obs}
\end{figure*}

A general analysis of orbital resonance and relativistic precession models within the context of tori has been very recently presented by \citet{Kotrlova2020}. They focus on sources in which a 3:2 (or some other) frequency ratio is observed (which constrains the value of r). Then their predictions are of the form  $Mf = F(a,\beta)$, where $\beta$ is proportional to the thickness of the torus ($\beta$ = 0 corresponds to a geodesic particle orbit).  Essentially all of their well-studied torus models produce results for $Mf$ which are comparable to or greater than that of the epicyclic orbital resonance model (plotted in Figures~\ref{fig:xrb_obs} and \ref{fig:smbh_obs}). A problem may arise for models such as these, since time dependent properties of the source, such as $\beta$, would produce a time dependence of the frequencies. For these reasons, we do not include these models in our figures. The hotspot model is also not included, since the predicted frequency depends on the radial location of the hotspot, which is unconstrained.

The predictions in the phase space of black hole mass, spin, and QPO frequency of the resonance, diskoseismic $g$-mode, non-axisymmetric $p$-modes, and disk-jet coupling models are plotted in Figures~\ref{fig:xrb_obs}~and~\ref{fig:smbh_obs}, where we also plot the orbital frequency of the ISCO. We do not include the warp model, since the plot of its lowest frequency is very close to that of the g-mode.


\section{Observations}
\label{sec:observations}

We focus here on detections of high-frequency QPOs in both X-ray binary and supermassive black hole systems. This neglects definite low-frequency QPOs and the kHz QPOs observed in neutron stars, as these exhibit distinct qualities and are likely to be due to other mechanisms.

\subsection{Stellar Mass Black Holes}
\label{sec:stellar_obs}

In Table~\ref{t:xrb_obs} we list observations of HFQPOs in the literature in black hole X-ray binaries with detection significance above 3$\sigma$ and for which masses and spins have been determined, excluding any sources that have been formally refuted or retracted since they were first reported. Most observations were taken with \emph{Rossi} X-ray Timing Explorer (RXTE), launded in 1995 and decommissioned in 2012.

In Figure~\ref{fig:xrb_obs}, we plot these observations. We show the stellar mass black holes both on the same scale as the upcoming supermassive plot, for easier comparison, and on a scale more appropriate for their own values. 

 The observed frequencies of HFQPOs do not shift freely with large luminosity changes \citep{Remillard2006}. The quality factor $Q = f / \mathrm{FWHM}$ of these HFQPOs range between 2 and 15, with fractional rms ranging from $\sim 1-7$\% \citep{Belloni2012}.

Two of the plotted sources exhibit an approximate 3:2 ratio in frequency of the HFQPO peaks: XTE~J1550-564 (184, 276 Hz) and GRO~J1655-40 (300, 450 Hz). GRS 1915+105 also has several detected HFQPOs that are not in a 3:2 ratio (41, 67, 166 Hz). While the 3:2 pairs have appeared simultaneously, most often they are detected singly, in separate observations \citep{Remillard2006}. In Figure~\ref{fig:xrb_obs} only the lower frequency of the pair is plotted for clarity; in the case of GRS~1915+105, all three QPOs are plotted, since no 3:2 harmonic ratios are present.

We note that HFQPOs in a 3:2 ratio have also been discovered in H~1743-322 \citep{Homan2005,Remillard2006}, but the mass estimates for this black hole are model-dependent and rely upon the QPO frequency itself. Such mass estimates cannot be used to constrain the various models without circular reasoning, so this object is excluded from our analysis. We further note that the definite 250~Hz QPO in XTE~J1650-500 was reported along with two additional marginal detections at $\sim110$ and $\sim170$~Hz that form an approximate 1:2:3 harmonic set with the 250~Hz detection. Because these detections are not highly significant and the harmonic ratio is speculative, we include only the 250~Hz oscillation in our analysis. Inclusion of the lower-frequency oscillations in Figure~\ref{fig:xrb_obs} would not change the conclusions of this paper, only shifting the green data point downwards by an amount encompassed by the measurement error.

Objects may appear more than once due to different mass or spin measurements in the literature. Spin is determined by one of two methods: thermal continuum fitting and iron-line spectroscopy, see Table~\ref{t:xrb_obs}. The continuum fitting method effectively determines the spin by fitting the thermal spectrum with a fully relativistic accretion disk model in a Kerr spacetime \citep[e.g.,][]{Shafee2006}. This is only possible when many of the other parameters in the fit, such as the black hole mass, inclination, and distance to the source are constrained by independent means. The iron-line spectroscopic method models the Fe K$\alpha$ fluorescence line generated through the irradiation of the disk by hard X-rays \citep[e.g.,][]{Brenneman2006}. The extent of the line's redshifted wing increases as the ISCO approaches the event horizon (i.e., with increasing spin, see Figure~\ref{fig:orbitals}). 

We note that in the case of GRS~1915+105, the maximal spin that we report here is found by both the reflection and continuum modeling methods in separate studies \citep{McClintock2006b,Miller2013}. However, \citet{Blum2009}, while finding the same maximal spin value if using the entire broadband spectrum, finds a much lower spin of $a\sim0.56$ if only the soft X-ray spectrum is used in the modeling.

\subsection{Supermassive Black Holes}
\label{sec:smbh_obs}

In Table~\ref{t:smbh_obs} we list observations of HFQPOs in the literature in AGN and tidal disruption events with detection significance above 3$\sigma$ and for which a reasonable mass estimate has been determined. Note that some HFQPOs have been detected in sources where there is no mass or spin determination. 

The identification of these oscillations as HF- rather than LFQPOs is achieved by several factors, which are typically the same from paper to paper and are based on similarities with better-determined objects. The first AGN QPO detection in RE~J1034+396 \citep{Gierlinski2008} has a high coherence and a measured black hole mass consistent with the known mass-frequency scaling relation for HFQPOs in XRBs. As in several of the other cases, assuming the detected QPO frequency was an LFQPO results in mass upper limits that are inconsistent with $M_\mathrm{BH}$ measurements from independent sources, and are often unfeasibly low for AGN ($\sim10^5$~M$_\odot$).  Additionally, the high accretion rate and strong soft excess in the X-ray energy spectrum and the shape of the broad-band noise component in the temporal power spectrum of all of the other claimed detections are similar to RE~J1034+396, while a number of the other papers cite also the high coherence and low fractional variability of their QPOs as consistent with those observed in BHB HFQPOs. In short, each paper makes the case for their detection being  HFQPO, which is our criterion for including them.

There are two AGN in which candidate low-frequency QPOs have been claimed: 2XMM~J1231+1106 \citep{Lin2013} and KIC~9650712 \citep{Smith2018b}. The LFQPO identification was based on mass scaling of frequency assuming a linear relationship between LFQPO periods and black hole mass \citep[see for example the Introduction of ][]{Carpano2018}, yielding consistent masses with other independent estimates, as well as low coherence and high fractional variability consistent with XRB LFQPOs. Because LFQPOs are not likely to be caused by the same mechanism as HFQPOs, we do not include them in this analysis.

Finally, we note that three QPOs have been claimed in extreme-UV light curves of AGN, which may have different physical origins and for which the authors do not explicitly speculate on the low- or high-frequency identity, although they do hypothesize that their origin is diskoseismic: TON~S~180, 1H~0419-577, and RXJ~0437.4-4711 \citep{Halpern2003}. Because these are the only detections not in X-rays and with no explicit frequency identification, we use different symbols in Figure~\ref{fig:smbh_obs}.

In Figure~\ref{fig:smbh_obs}, we plot these observations. In the case that spin measurements are unknown, objects are plotted as colored bands that account for the error in the mass and frequency measurements (where the error in the black hole mass overwhelmingly dominates). Frequencies have been corrected for cosmological redshift, and are given in the galaxy's rest frame. 

Although we only plot positive spins in Figure~\ref{fig:smbh_obs}, since all current spin observations are in this regime, it is important to remember that for supermassive black holes, retrograde spin ($a<0$) is a physically valid possibility. We therefore indicate the prediction for each of the theories for $a = -1$ on the left edge of the plot, for the curious. The spins reported here for SMBHs were measured using the Fe~K$\alpha$ line method, often modelled in tandem with X-ray spectroscopic fits; the details for each measurement can be found in the spin references in Table~\ref{t:smbh_obs}. Although thermal continuum fitting alone can in principle be used to determine the spin in AGN, it is not as straightforward, due mainly to the fact that the thermal disk continuum peaks in the extreme UV rather than the X-ray band, but also other complexities; see \citet{Reynolds2019} for a helpful summary. As is the case with stellar mass black holes, many of the other parameters of the fit including mass and accretion rate are constrained by independent observations, such as the width of broad optical Balmer lines \citep[e.g., ][]{Trakhtenbrot2012}  or the correlation between $M_\mathrm{BH}$ and the stellar velocity dispersion of the bulge \citep[$M-\sigma_*$, ][]{Ferrarese2012}. 

It is important to put the search for these SMBH QPOs into context. The variability of AGN is a red-noise process, and as such, can easily give rise to spurious periodic signals in both X-ray and UV/optical light curves \citep{Vaughan2005,Vaughan2016}. Because of this, many QPOs reported early on, typically before 2005, have been ruled out either by more careful analyses or by longer subsequent light curves. For our analysis, because our purpose is to establish whether any observations are consistent with the same models as employed for stellar-mass black holes, we include all AGN QPO detections that have not been formally retracted or refuted. All objects included in our plot have a detection significance in excess of 3$\sigma$ in their discovery papers. However, we note that how that significance is calculated varies from paper to paper; often, the significance is calculated only for regions of the light curve where the QPO is deemed present, while the significance of the QPO in the full light curve is much lower. The quality factors $Q$ of the QPOs are typically similar for AGN and stellar mass black holes, ranging from $\sim2-15$. The rms fractional variability amplitudes span a larger range and reach higher values than in stellar mass black holes, ranging from $2\% - 50\%$. The details for each object can be found by reading the QPO discovery references given in Table~\ref{t:smbh_obs}. 

Detection of two QPOs in small integer ratios is far less common in supermassive black holes than in stellar mass black holes. Although their detection is far from ubiquitous among XRBs, being seen in only three objects \citep{Motta2016}, there may not yet be a convincing case in AGN. When \citet{Zhang2017} reported the discovery of their 1.55$\times10^{-4}$~Hz QPO in Mrk~766, they noted that it would be in a 3:2 ratio with the previously-reported 2.4$\times10^{-4}$~Hz QPO in that same object \citep{Boller2001}, although the two peaks were not present at the same time. However, the \citet{Boller2001} result is more likely to be caused by red noise than an actual periodicity \citep{Vaughan2005}. The QPOs in ESO~113-G010 are in a 2:1 ratio \citep{Zhang2020}, and \citet{Alston2015} report the detection of a coherent soft lag in MS~2254.9–3712 that may indicate a harmonic component consistent with a 3:2 ratio. While such ratios may be consistent with resonance models generally, the ratios are approximate and the number of detections remains few.
\\
\\

\renewcommand{\arraystretch}{2}
\begin{deluxetable*}{lcccccc}

 \tablecaption{Observations of HFQPOs around Stellar Black Holes\label{t:xrb_obs}}
 \tablehead{
 \colhead{Name} &
   \colhead{$M_\mathrm{BH}$} &
  \colhead{BH Spin Range} &
  \colhead{$f_\mathrm{QPO}$ } &
  \colhead{$M_\mathrm{BH}$ Ref.} &
  \colhead{Spin Ref.} &
  \colhead{QPO Ref.} \\
\colhead{} &
\colhead{$M_\odot$} & 
 \colhead{$a$} &
 \colhead{Hz} &
 \colhead{} &
 \colhead{} &
\colhead{} \vspace{-5mm} \\}

\startdata
XTE J1550-564	&	$9.1^{+0.6}_{-0.6}$	&	$0.29 < a < 0.62^{r,c}$	&	184	&	1	&	7	&	14	\\
	&		&	$0.75 < a < 0.77^{r}$	&	184	&		&	8	&		\\
\hline
XTE J1650-500	&	$5.0^{+2.0}_{-2.0}$	&	$0.78 < a < 0.8^{r}$	&	250	&	2	&	8	&	15	\\
\hline
GROJ1655-40	&	$5.9^{+0.8}_{-0.8}$	&	$0.65 < a < 0.75^{c}$	&	300	&	3,4	&	9	&	14,16	\\
	&		&	$0.9 < a < 0.998^{r}$	&	300	&		&	10	&		\\
	&		&	$0.97 < a < 0.99^{r}$	&	300	&		&	8	&		\\
\hline
GRS1915+105	&	$10.1^{+0.6}_{-0.6}$	&$0.97 < a < 0.99^{r,c}$	&	41	&	5	&	11,12,13	&	17	\\
	&	$12.5^{+1.9}_{-1.9}$	&	$0.97 < a < 0.99^{r,c}$	&	41	&	6	&	11,12,13	&	17	 \\
	&	$10.1^{+0.6}_{-0.6}$	&$0.97 < a < 0.99^{r,c}$	&	67	&	5	&	11,12,13	&	18	\\
	&	$12.5^{+1.9}_{-1.9}$	&	$0.97 < a < 0.99^{r,c}$	&	67	&	6	&	11,12,13	&	18	 \\
    &	$10.1^{+0.6}_{-0.6}$	&$0.97 < a < 0.99^{r,c}$	&	166	&	5	&	11,12,13	&	19	 \\
	&	$12.5^{+1.9}_{-1.9}$	&	$0.97 < a < 0.99^{r,c}$	&	166	&	6	&	11,12,13	&	~19	 \vspace{-5mm}\\

\enddata

 \tablecomments{Observational data and references for HFQPOs around stellar mass black holes from the literature, in order of increasing right ascension. In objects where multiple HFQPOs are reported in small integer ratios, only the lowest frequency is given. Superscripts on the spin ranges indicate the measurement methods: Fe~K$\alpha$ reflection spectroscopy ($r$) or continuum fitting ($c$). References are as follows: $^{1}$\citet{Orosz2011},$^{2}$\citet{Orosz2004},$^{3}$\citet{Beer2002},$^{4}$\citet{Shahbaz2003},${^5}$\citet{Steeghs2013},$^{6}$\citet{Reid2014}
$^{7}$\citet{Steiner2011},$^{8}$\citet{Miller2009},$^{9}$\citet{Shafee2006},$^{10}$\citet{Reis2009},$^{11}$\citet{Blum2009},$^{12}$\citet{Miller2013},$^{13}$\citet{McClintock2006b},
$^{14}$\citet{Remillard2002},$^{15}$\citet{Homan2003},$^{16}$\citet{Remillard1999},$^{17}$\citet{Strohmayer2001},$^{18}$\citet{Morgan1997},$^{19}$\citet{Belloni2006}.}

 \end{deluxetable*}

\renewcommand{\arraystretch}{2}
\begin{deluxetable*}{lcccccccc}

 \tablecaption{Observations of QPOs around Supermassive Black Holes\label{t:smbh_obs}}
 \tablehead{
 \colhead{Name} &
   \colhead{BH Spin} &
  \colhead{log $M_\mathrm{BH}$} &
  \colhead{$f_\mathrm{QPO}$ } &
  \colhead{QPO Band} &
  \colhead{Object Type} &
  \colhead{QPO Ref.} &
  \colhead{$M_\mathrm{BH}$ Ref.} &
  \colhead{Spin Ref.} \\
\colhead{} &
\colhead{$a$} & 
 \colhead{$M_\odot$} &
 \colhead{Hz} &
 \colhead{} &
 \colhead{} &
 \colhead{} &
 \colhead{} &
\colhead{} \vspace{-5mm}\\}

\startdata
TON S 180	&	$< 0.4$	&	$6.85^{+0.5}_{-0.5}$	&	$5.56\times10^{-6}$	&	EUV	&	NLS1	&	1	&	11	&	21	\\
ESO 113-G010	&	0.998	&	$6.85^{+0.15}_{-0.24}$	&	$1.24\times10^{-4}$	&	X	&	NLS1	&	2	&	12	&	12	\\
ESO 113-G010	&	0.998	&	$6.85^{+0.15}_{-0.24}$	&	$6.79\times10^{-5}$	&	X	& NLS1	&	2	&	12	&	12	\\
1H0419-577	&	$>0.98$	&	$8.11^{+0.50}_{-0.50}$	&	$2.0\times10^{-6}$	&	EUV	&	Sy1	&	1	&	13	&	22	\\
RXJ 0437.4-4711	&	-	&	$7.77^{+0.5}_{-0.5}$	&	$1.27\times10^{-5}$	&	EUV	&	Sy1 &	1	&	13	&	-	\\
1H0707-495	&	$>0.976$	&	$6.36^{+0.24}_{-0.06}$	&	$2.6\times10^{-4}$	&	X	&	NLS1	&	3	&	14	&	23	\\
RE J1034+396	&	0.998	&	$6.0^{+1.0}_{-3.49}$	&	$2.7\times10^{-4}$	&	X	&	NLS1	&	4	&	15	&	15	\\
Mrk 766	&	$>0.92$	&	$6.82^{+0.05}_{-0.06}$	&	$1.55\times10^{-4}$	&	X	&	NLS1	&	5	&	16	&	24	\\
ASASSN-14li	&	$>0.7$	&	$6.23^{+0.35}_{-0.35}$	&	$7.7\times10^{-3}$	&	X	&	TDE	&	6	&	17	&	6	\\
MCG-06-30-15	&	$>0.917$	&	$6.20^{+0.09}_{-0.12}$	&	$2.73\times10^{-4}$	&	X	&	NLS1	&	7	&	18	&	25	\\
XMMU J134736.6+173403	&	-	&	$6.99^{+0.46}_{-0.20}$	&	$1.16\times10^{-5}$	&	X	&	Sy2	&	8	&	8	&	-	\\
Sw J164449.3+573451	&	-	&	$7.0^{+0.30}_{-0.35}$	&	$5.01\times10^{-3}$	&	X	&	TDE	&	9	&	19,20	&	-	\\
MS 2254.9-3712	&	-	&	$6.6^{+0.39}_{-0.60}$	&	$1.5\times10^{-4}$	&	X	&	NLS1	&	10	&	13	&	-	\vspace{-5mm}\\
 \enddata
 
 \tablecomments{Observational data and references for QPOs around supermassive black holes from the literature, in order of increasing right ascension. References are as follows: $^{1}$\citet{Halpern2003}, $^{2}$\citet{Zhang2020},$^{3}$\citet{Pan2016},$^{4}$\citet{Gierlinski2008},$^{5}$\citet{Zhang2017},$^{6}$\citet{Pasham2019},$^{7}$\citet{Gupta2018},$^{8}$\citet{Carpano2018}
$^{9}$\citet{Reis2012},$^{10}$\citet{Alston2015},$^{11}$\citet{Mathur2012},$^{12}$\citet{Cackett2013},$^{13}$\citet{Grupe2010},$^{14}$\citet{Zhou2005},$^{15}$\citet{Czerny2016},$^{16}$\citet{Bentz2015},$^{17}$\citet{vanVelzen2016},$^{18}$\citet{Bentz2016},$^{19}$\citet{Miller2011},$^{20}$\citet{Seifina2017},$^{21}$\citet{Parker2018},$^{22}$\citet{Jiang2019},$^{23}$\citet{Zoghbi2010},$^{24}$\citet{Buisson2018},$^{25}$\citet{Miniutti2007}.}

 \end{deluxetable*}

\section{Discussion} 
\label{sec:discussion}

It is clear from Figures~\ref{fig:xrb_obs} and \ref{fig:smbh_obs} that more theoretical models are consistent with observations of HFQPOs around stellar mass black holes than those around supermassive black holes. Although no single physical model explains all of the stellar mass observations, they share generally the same parameter space. The only exception is the lowest frequency HFQPO candidate in GRS~1915+105 \citep[41~Hz; ][]{Strohmayer2001}, which falls among the SMBH values of $Mf$; this QPO was not detected in any follow-up studies, however.

For HFQPOs in AGN, none of the models we tested are consistent with the observations.

The HFQPOs in AGN \emph{are} consistent with axisymmetric pressure modes, rather than the gravity modes, in diskoseismic models \citep[e.g.,][]{Wagoner2001,Ortega2002}. However, as noted in Section~\ref{sec:diskoseismology}, their very small radial extent and tendency to leakage through the ISCO should preclude their observation.

For tidal disruption events, the situation is different: the two cases (Swift~J1644+57 and ASAS-SN~14li) are consistent with each other, and have higher values of $M/10M_\odot \times f_\mathrm{QPO}$ than the AGN or stellar mass black holes, qualitatively consistent with the orbital frequency at the ISCO. This was the model favored by \citet{Pasham2019} in the discovery paper for the QPO in ASAS-SN~14li; they also point out that this object has the highest dimensionless frequency of any black hole QPO discovered thus far.

Discoveries of QPOs have also been reported in two candidate intermediate-mass black holes: NGC~5408~X1 \citep{Strohmayer2009} and M82~X1 \citep{Dewangan2006,Pasham2014}. In both cases, however, the reported black hole masses are derived from the observed QPO frequency, and not an independent method (which is very difficult indeed for intermediate mass black holes), so we cannot include them in our analysis. We note, however, that both of these objects have higher quality factors ($Q\sim20-40$) than QPOs reported in AGN, TDEs, or XRBs. 

Naively, it appears that HFQPOs in AGN and stellar mass black holes must be caused by different physical conditions. This has serious ramifications for the scale invariance of accretion. However, it is possible that the same underlying phenomenon is still responsible for both if the governing theory contains an additional dimensionless parameter, such as $L/L_\mathrm{Edd}$.

Some insight into the relative origins of HFQPOs in each object class can be garnered from comparing the ``states" in which they occur. As mentioned in the introduction, HFQPOs in XRBs apparently occur in high-luminosity, possibly high-accretion rate states with a soft X-ray spectrum containing a significant thermal disk component and a steep power law component. These states are often referred to as the steep power law state or the soft-intermediate state. In AGN, the accretion disk radiates in the UV and optical, and not the X-rays; X-ray emission comes from the coronal plasma located above the black hole or inner disk regions where electrons Compton-upscatter accretion disk photons to high energies.  Most SMBH QPOs have been discovered in a class of AGN known as Narrow Line Seyfert~1 (NLS1) galaxies, characterised by broad Balmer lines with unusually low velocity widths, significant Fe~II emission, and a very soft X-ray spectrum \citep[e.g.,][]{Osterbrock1985,Komossa2008}. It has been pointed out that NLS1s are possibly analogous to the ``high-soft" state X-ray binaries \citep{Pounds1995,Boller1996}; however, this was before the nomenclature became richer in later studies, differentiating the purely thermal, high-soft state from the intermediate states in which HFQPOs are frequently seen; see \citet{Klein2008} for an overview. Indeed, NLS1s are qualitatively similar to the intermediate XRB states in many ways beyond the X-ray spectral components: the NLS1 class also has higher-amplitude rapid X-ray variability compared to other Seyfert~1 AGN, also a feature of the ``anomalous" state in which HFQPOs are seen in XRBs \citep{Belloni2012}, and tend to have higher accretion rates and lower black hole masses than typical Seyfert~1 AGN \citep{Williams2018}. 

There is, however, a potentially major selection effect at play. In order to significantly detect a QPO over many cycles in an AGN, an extraordinarily long X-ray observation is needed. All AGN QPOs have been discovered in archival observations that were taken for other purposes, usually to study the Fe-K$\alpha$ line, reverberation, or other spectral phenomena. Because of their relatively high luminosities and high-amplitude X-ray variability, NLS1s have been favored targets for such observations. It is not yet known whether light curves of similar length and quality for a similar population of ordinary Seyfert~1 galaxies would yield QPOs at the same rate, especially since the duty cycle of AGN QPOs is unknown. It is worth noting, however, that \citet{Gonzalez2012} found that out of 104 AGN with archival XMM-\emph{Newton} observations, only RE~J1034+396 showed clear evidence of a QPO. Several other AGN considered in this paper are among their sample, with QPOs either discovered more recently, with different methods, or in different wavebands.

\section{Conclusion}
\label{sec:conclusion}

We have collected all observations of HFQPOs in the literature discovered in black hole X-ray binaries, tidal disruption events, and AGN. These observations are compared to theoretical models of QPOs that make quantitative predictions about the observed frequences: orbital resonance models, diskoseismic $g$-modes, non-axisymmetric $p$-modes, and disk-jet coupling models, as well as the orbital frequency at the ISCO. We find that while most HFQPOs in black hole XRBs tend to be well-described by at least one such model, HFQPOs in AGN are not. Additionally, the QPOs in AGN appear to be different from those in TDEs, which occur at approximately the ISCO orbital frequency. We propose that the mechanism responsible for QPOs must depend upon more than the mass and spin of the black hole, perhaps an additional dimensionless parameter like the Eddington ratio, despite an observed correlation of mass and QPO frequency across many decades in mass. Otherwise, the QPOs must be generated by different physical mechanisms, complicating the paradigm of accretion scale-invariance.

\acknowledgments
Support for KLS was provided by the National Aeronautics and
Space Administration through Einstein Postdoctoral Fellowship
Award Number PF7-180168, issued by the Chandra
X-ray Observatory Center, which is operated by the
Smithsonian Astrophysical Observatory for and on behalf
of the National Aeronautics Space Administration
under contract NAS8-03060. CT acknowledges financial support from the Stanford Physics Department Summer Research Program.

\small
\bibliographystyle{aasjournal}
\bibliography{qpo_paper}

\begin{thebibliography}{}
\expandafter\ifx\csname natexlab\endcsname\relax\def\natexlab#1{#1}\fi
\providecommand{\url}[1]{\href{#1}{#1}}
\providecommand{\dodoi}[1]{doi:~\href{http://doi.org/#1}{\nolinkurl{#1}}}
\providecommand{\doeprint}[1]{\href{http://ascl.net/#1}{\nolinkurl{http://ascl.net/#1}}}
\providecommand{\doarXiv}[1]{\href{https://arxiv.org/abs/#1}{\nolinkurl{https://arxiv.org/abs/#1}}}

\bibitem[{{Abramowicz} {et~al.}(2003){Abramowicz}, {Karas}, {Kluzniak}, {Lee},
  \& {Rebusco}}]{Abramowicz2003}
{Abramowicz}, M.~A., {Karas}, V., {Kluzniak}, W., {Lee}, W.~H., \& {Rebusco},
  P. 2003, \pasj, 55, 467, \dodoi{10.1093/pasj/55.2.467}

\bibitem[{{Abramowicz} \& {Klu{\'z}niak}(2001)}]{Abramowicz2001}
{Abramowicz}, M.~A., \& {Klu{\'z}niak}, W. 2001, \aap, 374, L19,
  \dodoi{10.1051/0004-6361:20010791}

\bibitem[{{Abramowicz} {et~al.}(2004){Abramowicz}, {Klu{\'z}niak},
  {McClintock}, \& {Remillard}}]{Abramowicz2004}
{Abramowicz}, M.~A., {Klu{\'z}niak}, W., {McClintock}, J.~E., \& {Remillard},
  R.~A. 2004, \apjl, 609, L63, \dodoi{10.1086/422810}

\bibitem[{{Alston} {et~al.}(2015){Alston}, {Parker},
  {Markevi{\v{c}}i{\={u}}t{\.{e}}}, {Fabian}, {Middleton}, {Lohfink}, {Kara},
  \& {Pinto}}]{Alston2015}
{Alston}, W.~N., {Parker}, M.~L., {Markevi{\v{c}}i{\={u}}t{\.{e}}}, J.,
  {et~al.} 2015, \mnras, 449, 467, \dodoi{10.1093/mnras/stv351}

\bibitem[{{Beer} \& {Podsiadlowski}(2002)}]{Beer2002}
{Beer}, M.~E., \& {Podsiadlowski}, P. 2002, \mnras, 331, 351,
  \dodoi{10.1046/j.1365-8711.2002.05189.x}

\bibitem[{{Belloni} {et~al.}(2006){Belloni}, {Soleri}, {Casella}, {M{\'e}ndez},
  \& {Migliari}}]{Belloni2006}
{Belloni}, T., {Soleri}, P., {Casella}, P., {M{\'e}ndez}, M., \& {Migliari}, S.
  2006, \mnras, 369, 305, \dodoi{10.1111/j.1365-2966.2006.10286.x}

\bibitem[{{Belloni} {et~al.}(2019){Belloni}, {Bhattacharya}, {Caccese},
  {Bhalerao}, {Vadawale}, \& {Yadav}}]{Belloni2019}
{Belloni}, T.~M., {Bhattacharya}, D., {Caccese}, P., {et~al.} 2019, \mnras,
  489, 1037, \dodoi{10.1093/mnras/stz2143}

\bibitem[{{Belloni} {et~al.}(2012){Belloni}, {Sanna}, \&
  {M{\'e}ndez}}]{Belloni2012}
{Belloni}, T.~M., {Sanna}, A., \& {M{\'e}ndez}, M. 2012, \mnras, 426, 1701,
  \dodoi{10.1111/j.1365-2966.2012.21634.x}

\bibitem[{{Bentz} {et~al.}(2016){Bentz}, {Cackett}, {Crenshaw}, {Horne},
  {Street}, \& {Ou-Yang}}]{Bentz2016}
{Bentz}, M.~C., {Cackett}, E.~M., {Crenshaw}, D.~M., {et~al.} 2016, \apj, 830,
  136, \dodoi{10.3847/0004-637X/830/2/136}

\bibitem[{{Bentz} \& {Katz}(2015)}]{Bentz2015}
{Bentz}, M.~C., \& {Katz}, S. 2015, \pasp, 127, 67, \dodoi{10.1086/679601}

\bibitem[{{Blum} {et~al.}(2009){Blum}, {Miller}, {Fabian}, {Miller}, {Homan},
  {van der Klis}, {Cackett}, \& {Reis}}]{Blum2009}
{Blum}, J.~L., {Miller}, J.~M., {Fabian}, A.~C., {et~al.} 2009, \apj, 706, 60,
  \dodoi{10.1088/0004-637X/706/1/60}

\bibitem[{{Boller} {et~al.}(1996){Boller}, {Brandt}, \& {Fink}}]{Boller1996}
{Boller}, T., {Brandt}, W.~N., \& {Fink}, H. 1996, \aap, 305, 53.
\newblock \doarXiv{astro-ph/9504093}

\bibitem[{{Boller} {et~al.}(2001){Boller}, {Keil}, {Tr{\"u}mper}, {O'Brien},
  {Reeves}, \& {Page}}]{Boller2001}
{Boller}, T., {Keil}, R., {Tr{\"u}mper}, J., {et~al.} 2001, \aap, 365, L146,
  \dodoi{10.1051/0004-6361:20000083}

\bibitem[{{Brenneman} \& {Reynolds}(2006)}]{Brenneman2006}
{Brenneman}, L.~W., \& {Reynolds}, C.~S. 2006, \apj, 652, 1028,
  \dodoi{10.1086/508146}

\bibitem[{{Buisson} {et~al.}(2018){Buisson}, {Parker}, {Kara}, {Vasudevan},
  {Lohfink}, {Pinto}, {Fabian}, {Ballantyne}, {Boggs}, {Christensen}, {Craig},
  {Farrah}, {Hailey}, {Harrison}, {Ricci}, {Stern}, {Walton}, \&
  {Zhang}}]{Buisson2018}
{Buisson}, D.~J.~K., {Parker}, M.~L., {Kara}, E., {et~al.} 2018, \mnras, 480,
  3689, \dodoi{10.1093/mnras/sty2081}

\bibitem[{{Cackett} {et~al.}(2013){Cackett}, {Fabian}, {Zogbhi}, {Kara},
  {Reynolds}, \& {Uttley}}]{Cackett2013}
{Cackett}, E.~M., {Fabian}, A.~C., {Zogbhi}, A., {et~al.} 2013, \apjl, 764, L9,
  \dodoi{10.1088/2041-8205/764/1/L9}

\bibitem[{{Carpano} \& {Jin}(2018)}]{Carpano2018}
{Carpano}, S., \& {Jin}, C. 2018, \mnras, 477, 3178,
  \dodoi{10.1093/mnras/sty841}

\bibitem[{{Czerny} {et~al.}(2016){Czerny}, {You}, {Kurcz},
  {{\'S}redzi{\'n}ska}, {Hryniewicz}, {Niko{\l}ajuk}, {Krupa}, {Wang}, {Hu}, \&
  {{\.Z}ycki}}]{Czerny2016}
{Czerny}, B., {You}, B., {Kurcz}, A., {et~al.} 2016, \aap, 594, A102,
  \dodoi{10.1051/0004-6361/201628103}

\bibitem[{{Dewangan} {et~al.}(2006){Dewangan}, {Titarchuk}, \&
  {Griffiths}}]{Dewangan2006}
{Dewangan}, G.~C., {Titarchuk}, L., \& {Griffiths}, R.~E. 2006, \apjl, 637,
  L21, \dodoi{10.1086/499235}

\bibitem[{{Dewberry} {et~al.}(2020){Dewberry}, {Latter}, {Ogilvie}, \&
  {Fromang}}]{Dewberry2020}
{Dewberry}, J.~W., {Latter}, H.~N., {Ogilvie}, G.~I., \& {Fromang}, S. 2020,
  arXiv e-prints, arXiv:2006.16266.
\newblock \doarXiv{2006.16266}

\bibitem[{{Ferrarese} \& {Merritt}(2000)}]{Ferrarese2012}
{Ferrarese}, L., \& {Merritt}, D. 2000, \apjl, 539, L9, \dodoi{10.1086/312838}

\bibitem[{{Gierli{\'n}ski} {et~al.}(2008){Gierli{\'n}ski}, {Middleton}, {Ward},
  \& {Done}}]{Gierlinski2008}
{Gierli{\'n}ski}, M., {Middleton}, M., {Ward}, M., \& {Done}, C. 2008, \nat,
  455, 369, \dodoi{10.1038/nature07277}

\bibitem[{{Gonz{\'a}lez-Mart{\'\i}n} \& {Vaughan}(2012)}]{Gonzalez2012}
{Gonz{\'a}lez-Mart{\'\i}n}, O., \& {Vaughan}, S. 2012, \aap, 544, A80,
  \dodoi{10.1051/0004-6361/201219008}

\bibitem[{{Grupe} {et~al.}(2010){Grupe}, {Komossa}, {Leighly}, \&
  {Page}}]{Grupe2010}
{Grupe}, D., {Komossa}, S., {Leighly}, K.~M., \& {Page}, K.~L. 2010, \apjs,
  187, 64, \dodoi{10.1088/0067-0049/187/1/64}

\bibitem[{{Gupta} {et~al.}(2018){Gupta}, {Tripathi}, {Wiita}, {Gu}, {Bambi}, \&
  {Ho}}]{Gupta2018}
{Gupta}, A.~C., {Tripathi}, A., {Wiita}, P.~J., {et~al.} 2018, \aap, 616, L6,
  \dodoi{10.1051/0004-6361/201833629}

\bibitem[{{Halpern} {et~al.}(2003){Halpern}, {Leighly}, \&
  {Marshall}}]{Halpern2003}
{Halpern}, J.~P., {Leighly}, K.~M., \& {Marshall}, H.~L. 2003, \apj, 585, 665,
  \dodoi{10.1086/346106}

\bibitem[{{Homan} {et~al.}(2003){Homan}, {Klein-Wolt}, {Rossi}, {Miller},
  {Wijnands}, {Belloni}, {van der Klis}, \& {Lewin}}]{Homan2003}
{Homan}, J., {Klein-Wolt}, M., {Rossi}, S., {et~al.} 2003, \apj, 586, 1262,
  \dodoi{10.1086/367699}

\bibitem[{{Homan} {et~al.}(2005){Homan}, {Miller}, {Wijnands}, {van der Klis},
  {Belloni}, {Steeghs}, \& {Lewin}}]{Homan2005}
{Homan}, J., {Miller}, J.~M., {Wijnands}, R., {et~al.} 2005, \apj, 623, 383,
  \dodoi{10.1086/424994}

\bibitem[{{Hor{\'a}k}(2008)}]{Horak2008}
{Hor{\'a}k}, J. 2008, \aap, 486, 1, \dodoi{10.1051/0004-6361:20078305}

\bibitem[{{Jiang} {et~al.}(2019){Jiang}, {Walton}, {Fabian}, \&
  {Parker}}]{Jiang2019}
{Jiang}, J., {Walton}, D.~J., {Fabian}, A.~C., \& {Parker}, M.~L. 2019, \mnras,
  483, 2958, \dodoi{10.1093/mnras/sty3228}

\bibitem[{{Kato} \& {Fukue}(1980)}]{Kato1980}
{Kato}, S., \& {Fukue}, J. 1980, \pasj, 32, 377

\bibitem[{{Kato} {et~al.}(2008){Kato}, {Fukue}, \& {Mineshige}}]{Kato2008}
{Kato}, S., {Fukue}, J., \& {Mineshige}, S. 2008, {Black-Hole Accretion Disks
  --- Towards a New Paradigm ---}

\bibitem[{{Klein-Wolt} \& {van der Klis}(2008)}]{Klein2008}
{Klein-Wolt}, M., \& {van der Klis}, M. 2008, \apj, 675, 1407,
  \dodoi{10.1086/525843}

\bibitem[{{Komossa}(2008)}]{Komossa2008}
{Komossa}, S. 2008, in Revista Mexicana de Astronomia y Astrofisica Conference
  Series, Vol.~32, Revista Mexicana de Astronomia y Astrofisica Conference
  Series, 86--92

\bibitem[{{Kotrlov{\'a}} {et~al.}(2020){Kotrlov{\'a}},
  {{\v{S}}r{\'a}mkov{\'a}}, {T{\"o}r{\"o}k}, {Goluchov{\'a}},
  {Ji{\v{r}}{\'\i}}, {Straub}, {Lan{\v{c}}ov{\'a}}, {Stuchl{\'\i}k}, \&
  {Abramowicz}}]{Kotrlova2020}
{Kotrlov{\'a}}, A., {{\v{S}}r{\'a}mkov{\'a}}, E., {T{\"o}r{\"o}k}, G., {et~al.}
  2020, arXiv e-prints, arXiv:2008.12963.
\newblock \doarXiv{2008.12963}

\bibitem[{{Lai} {et~al.}(2013){Lai}, {Fu}, {Tsang}, {Horak}, \& {Yu}}]{Lai2013}
{Lai}, D., {Fu}, W., {Tsang}, D., {Horak}, J., \& {Yu}, C. 2013, in IAU
  Symposium, Vol. 290, Feeding Compact Objects: Accretion on All Scales, ed.
  C.~M. {Zhang}, T.~{Belloni}, M.~{M{\'e}ndez}, \& S.~N. {Zhang}, 57--61

\bibitem[{{Li} \& {Narayan}(2004)}]{Li2004}
{Li}, L.-X., \& {Narayan}, R. 2004, \apj, 601, 414, \dodoi{10.1086/380446}

\bibitem[{{Lin} {et~al.}(2013){Lin}, {Irwin}, {Godet}, {Webb}, \&
  {Barret}}]{Lin2013}
{Lin}, D., {Irwin}, J.~A., {Godet}, O., {Webb}, N.~A., \& {Barret}, D. 2013,
  \apjl, 776, L10, \dodoi{10.1088/2041-8205/776/1/L10}

\bibitem[{{Mathur} {et~al.}(2012){Mathur}, {Fields}, {Peterson}, \&
  {Grupe}}]{Mathur2012}
{Mathur}, S., {Fields}, D., {Peterson}, B.~M., \& {Grupe}, D. 2012, \apj, 754,
  146, \dodoi{10.1088/0004-637X/754/2/146}

\bibitem[{{McClintock} {et~al.}(2006){McClintock}, {Shafee}, {Narayan},
  {Remillard}, {Davis}, \& {Li}}]{McClintock2006b}
{McClintock}, J.~E., {Shafee}, R., {Narayan}, R., {et~al.} 2006, \apj, 652,
  518, \dodoi{10.1086/508457}

\bibitem[{{McKinney} {et~al.}(2012){McKinney}, {Tchekhovskoy}, \&
  {Blandford}}]{McKinney2012}
{McKinney}, J.~C., {Tchekhovskoy}, A., \& {Blandford}, R.~D. 2012, \mnras, 423,
  3083, \dodoi{10.1111/j.1365-2966.2012.21074.x}

\bibitem[{{M{\"\i}ller} \& {G{\"u}ltekin}(2011)}]{Miller2011}
{M{\"\i}ller}, J.~M., \& {G{\"u}ltekin}, K. 2011, \apjl, 738, L13,
  \dodoi{10.1088/2041-8205/738/1/L13}

\bibitem[{{Miller} {et~al.}(2009){Miller}, {Reynolds}, {Fabian}, {Miniutti}, \&
  {Gallo}}]{Miller2009}
{Miller}, J.~M., {Reynolds}, C.~S., {Fabian}, A.~C., {Miniutti}, G., \&
  {Gallo}, L.~C. 2009, \apj, 697, 900, \dodoi{10.1088/0004-637X/697/1/900}

\bibitem[{{Miller} {et~al.}(2013){Miller}, {Parker}, {Fuerst}, {Bachetti},
  {Harrison}, {Barret}, {Boggs}, {Chakrabarty}, {Christensen}, {Craig},
  {Fabian}, {Grefenstette}, {Hailey}, {King}, {Stern}, {Tomsick}, {Walton}, \&
  {Zhang}}]{Miller2013}
{Miller}, J.~M., {Parker}, M.~L., {Fuerst}, F., {et~al.} 2013, \apjl, 775, L45,
  \dodoi{10.1088/2041-8205/775/2/L45}

\bibitem[{{Miniutti} {et~al.}(2007){Miniutti}, {Fabian}, {Anabuki}, {Crummy},
  {Fukazawa}, {Gallo}, {Haba}, {Hayashida}, {Holt}, {Kunieda}, {Larsson},
  {Markowitz}, {Matsumoto}, {Ohno}, {Reeves}, {Takahashi}, {Tanaka},
  {Terashima}, {Torii}, {Ueda}, {Ushio}, {Watanabe}, {Yamauchi}, \&
  {Yaqoob}}]{Miniutti2007}
{Miniutti}, G., {Fabian}, A.~C., {Anabuki}, N., {et~al.} 2007, \pasj, 59, 315,
  \dodoi{10.1093/pasj/59.sp1.S315}

\bibitem[{{Morgan} {et~al.}(1997){Morgan}, {Remillard}, \&
  {Greiner}}]{Morgan1997}
{Morgan}, E.~H., {Remillard}, R.~A., \& {Greiner}, J. 1997, \apj, 482, 993,
  \dodoi{10.1086/304191}

\bibitem[{{Motta}(2016)}]{Motta2016}
{Motta}, S.~E. 2016, Astronomische Nachrichten, 337, 398,
  \dodoi{10.1002/asna.201612320}

\bibitem[{{O'Neill} {et~al.}(2009){O'Neill}, {Reynolds}, \&
  {Miller}}]{Oneill2009}
{O'Neill}, S.~M., {Reynolds}, C.~S., \& {Miller}, M.~C. 2009, \apj, 693, 1100,
  \dodoi{10.1088/0004-637X/693/2/1100}

\bibitem[{{Orosz} {et~al.}(2004){Orosz}, {McClintock}, {Remillard}, \&
  {Corbel}}]{Orosz2004}
{Orosz}, J.~A., {McClintock}, J.~E., {Remillard}, R.~A., \& {Corbel}, S. 2004,
  \apj, 616, 376, \dodoi{10.1086/424892}

\bibitem[{{Orosz} {et~al.}(2011){Orosz}, {Steiner}, {McClintock}, {Torres},
  {Remillard}, {Bailyn}, \& {Miller}}]{Orosz2011}
{Orosz}, J.~A., {Steiner}, J.~F., {McClintock}, J.~E., {et~al.} 2011, \apj,
  730, 75, \dodoi{10.1088/0004-637X/730/2/75}

\bibitem[{{Ortega-Rodr{\'\i}guez} {et~al.}(2002){Ortega-Rodr{\'\i}guez},
  {Silbergleit}, \& {Wagoner}}]{Ortega2002}
{Ortega-Rodr{\'\i}guez}, M., {Silbergleit}, A.~S., \& {Wagoner}, R.~V. 2002,
  \apj, 567, 1043, \dodoi{10.1086/338685}

\bibitem[{{Osterbrock} \& {Pogge}(1985)}]{Osterbrock1985}
{Osterbrock}, D.~E., \& {Pogge}, R.~W. 1985, \apj, 297, 166,
  \dodoi{10.1086/163513}

\bibitem[{{Pan} {et~al.}(2016){Pan}, {Yuan}, {Yao}, {Zhou}, {Liu}, {Zhou}, \&
  {Zhang}}]{Pan2016}
{Pan}, H.-W., {Yuan}, W., {Yao}, S., {et~al.} 2016, \apjl, 819, L19,
  \dodoi{10.3847/2041-8205/819/2/L19}

\bibitem[{{Parker} {et~al.}(2018){Parker}, {Miller}, \& {Fabian}}]{Parker2018}
{Parker}, M.~L., {Miller}, J.~M., \& {Fabian}, A.~C. 2018, \mnras, 474, 1538,
  \dodoi{10.1093/mnras/stx2861}

\bibitem[{{Pasham} {et~al.}(2014){Pasham}, {Strohmayer}, \&
  {Mushotzky}}]{Pasham2014}
{Pasham}, D.~R., {Strohmayer}, T.~E., \& {Mushotzky}, R.~F. 2014, \nat, 513,
  74, \dodoi{10.1038/nature13710}

\bibitem[{{Pasham} {et~al.}(2019){Pasham}, {Remillard}, {Fragile}, {Franchini},
  {Stone}, {Lodato}, {Homan}, {Chakrabarty}, {Baganoff}, {Steiner}, {Coughlin},
  \& {Pasham}}]{Pasham2019}
{Pasham}, D.~R., {Remillard}, R.~A., {Fragile}, P.~C., {et~al.} 2019, Science,
  363, 531, \dodoi{10.1126/science.aar7480}

\bibitem[{{Perez} {et~al.}(1997){Perez}, {Silbergleit}, {Wagoner}, \&
  {Lehr}}]{Perez1997}
{Perez}, C.~A., {Silbergleit}, A.~S., {Wagoner}, R.~V., \& {Lehr}, D.~E. 1997,
  \apj, 476, 589, \dodoi{10.1086/303658}

\bibitem[{{Pounds} {et~al.}(1995){Pounds}, {Done}, \& {Osborne}}]{Pounds1995}
{Pounds}, K.~A., {Done}, C., \& {Osborne}, J.~P. 1995, \mnras, 277, L5,
  \dodoi{10.1093/mnras/277.1.L5}

\bibitem[{{Reid} {et~al.}(2014){Reid}, {McClintock}, {Steiner}, {Steeghs},
  {Remillard}, {Dhawan}, \& {Narayan}}]{Reid2014}
{Reid}, M.~J., {McClintock}, J.~E., {Steiner}, J.~F., {et~al.} 2014, \apj, 796,
  2, \dodoi{10.1088/0004-637X/796/1/2}

\bibitem[{{Reis} {et~al.}(2009){Reis}, {Fabian}, {Ross}, \&
  {Miller}}]{Reis2009}
{Reis}, R.~C., {Fabian}, A.~C., {Ross}, R.~R., \& {Miller}, J.~M. 2009, \mnras,
  395, 1257, \dodoi{10.1111/j.1365-2966.2009.14622.x}

\bibitem[{{Reis} {et~al.}(2012){Reis}, {Miller}, {Reynolds}, {G{\"u}ltekin},
  {Maitra}, {King}, \& {Strohmayer}}]{Reis2012}
{Reis}, R.~C., {Miller}, J.~M., {Reynolds}, M.~T., {et~al.} 2012, Science, 337,
  949, \dodoi{10.1126/science.1223940}

\bibitem[{{Remillard} \& {McClintock}(2006)}]{Remillard2006}
{Remillard}, R.~A., \& {McClintock}, J.~E. 2006, \araa, 44, 49,
  \dodoi{10.1146/annurev.astro.44.051905.092532}

\bibitem[{{Remillard} {et~al.}(1999){Remillard}, {Morgan}, {McClintock},
  {Bailyn}, \& {Orosz}}]{Remillard1999}
{Remillard}, R.~A., {Morgan}, E.~H., {McClintock}, J.~E., {Bailyn}, C.~D., \&
  {Orosz}, J.~A. 1999, \apj, 522, 397, \dodoi{10.1086/307606}

\bibitem[{{Remillard} {et~al.}(2002){Remillard}, {Muno}, {McClintock}, \&
  {Orosz}}]{Remillard2002}
{Remillard}, R.~A., {Muno}, M.~P., {McClintock}, J.~E., \& {Orosz}, J.~A. 2002,
  \apj, 580, 1030, \dodoi{10.1086/343791}

\bibitem[{{Reynolds}(2019)}]{Reynolds2019}
{Reynolds}, C.~S. 2019, Nature Astronomy, 3, 41,
  \dodoi{10.1038/s41550-018-0665-z}

\bibitem[{{Reynolds} \& {Miller}(2009)}]{Reynolds2009}
{Reynolds}, C.~S., \& {Miller}, M.~C. 2009, \apj, 692, 869,
  \dodoi{10.1088/0004-637X/692/1/869}

\bibitem[{{Rezzolla} {et~al.}(2003){Rezzolla}, {Yoshida}, {Maccarone}, \&
  {Zanotti}}]{Rezzolla2003}
{Rezzolla}, L., {Yoshida}, S., {Maccarone}, T.~J., \& {Zanotti}, O. 2003,
  \mnras, 344, L37, \dodoi{10.1046/j.1365-8711.2003.07018.x}

\bibitem[{{Seifina} {et~al.}(2017){Seifina}, {Titarchuk}, \&
  {Virgilli}}]{Seifina2017}
{Seifina}, E., {Titarchuk}, L., \& {Virgilli}, E. 2017, \aap, 607, A38,
  \dodoi{10.1051/0004-6361/201730869}

\bibitem[{{Shafee} {et~al.}(2006){Shafee}, {McClintock}, {Narayan}, {Davis},
  {Li}, \& {Remillard}}]{Shafee2006}
{Shafee}, R., {McClintock}, J.~E., {Narayan}, R., {et~al.} 2006, \apjl, 636,
  L113, \dodoi{10.1086/498938}

\bibitem[{{Shahbaz}(2003)}]{Shahbaz2003}
{Shahbaz}, T. 2003, \mnras, 339, 1031, \dodoi{10.1046/j.1365-8711.2003.06258.x}

\bibitem[{{Smith} {et~al.}(2018){Smith}, {Mushotzky}, {Boyd}, \&
  {Wagoner}}]{Smith2018b}
{Smith}, K.~L., {Mushotzky}, R.~F., {Boyd}, P.~T., \& {Wagoner}, R.~V. 2018,
  \apjl, 860, L10, \dodoi{10.3847/2041-8213/aac88c}

\bibitem[{{Steeghs} {et~al.}(2013){Steeghs}, {McClintock}, {Parsons}, {Reid},
  {Littlefair}, \& {Dhillon}}]{Steeghs2013}
{Steeghs}, D., {McClintock}, J.~E., {Parsons}, S.~G., {et~al.} 2013, \apj, 768,
  185, \dodoi{10.1088/0004-637X/768/2/185}

\bibitem[{{Steiner} {et~al.}(2011){Steiner}, {Reis}, {McClintock}, {Narayan},
  {Remillard}, {Orosz}, {Gou}, {Fabian}, \& {Torres}}]{Steiner2011}
{Steiner}, J.~F., {Reis}, R.~C., {McClintock}, J.~E., {et~al.} 2011, \mnras,
  416, 941, \dodoi{10.1111/j.1365-2966.2011.19089.x}

\bibitem[{{Stella} {et~al.}(1999){Stella}, {Vietri}, \& {Morsink}}]{Stella1999}
{Stella}, L., {Vietri}, M., \& {Morsink}, S.~M. 1999, \apjl, 524, L63,
  \dodoi{10.1086/312291}

\bibitem[{{Strohmayer}(2001)}]{Strohmayer2001}
{Strohmayer}, T.~E. 2001, \apjl, 554, L169, \dodoi{10.1086/321720}

\bibitem[{{Strohmayer} \& {Mushotzky}(2009)}]{Strohmayer2009}
{Strohmayer}, T.~E., \& {Mushotzky}, R.~F. 2009, \apj, 703, 1386,
  \dodoi{10.1088/0004-637X/703/2/1386}

\bibitem[{{Trakhtenbrot} \& {Netzer}(2012)}]{Trakhtenbrot2012}
{Trakhtenbrot}, B., \& {Netzer}, H. 2012, \mnras, 427, 3081,
  \dodoi{10.1111/j.1365-2966.2012.22056.x}

\bibitem[{{van Velzen} {et~al.}(2016){van Velzen}, {Anderson}, {Stone},
  {Fraser}, {Wevers}, {Metzger}, {Jonker}, {van der Horst}, {Staley}, {Mendez},
  {Miller-Jones}, {Hodgkin}, {Campbell}, \& {Fender}}]{vanVelzen2016}
{van Velzen}, S., {Anderson}, G.~E., {Stone}, N.~C., {et~al.} 2016, Science,
  351, 62, \dodoi{10.1126/science.aad1182}

\bibitem[{{Vaughan}(2005)}]{Vaughan2005}
{Vaughan}, S. 2005, \aap, 431, 391, \dodoi{10.1051/0004-6361:20041453}

\bibitem[{{Vaughan} {et~al.}(2016){Vaughan}, {Uttley}, {Markowitz},
  {Huppenkothen}, {Middleton}, {Alston}, {Scargle}, \& {Farr}}]{Vaughan2016}
{Vaughan}, S., {Uttley}, P., {Markowitz}, A.~G., {et~al.} 2016, \mnras, 461,
  3145, \dodoi{10.1093/mnras/stw1412}

\bibitem[{{Wagoner} {et~al.}(2001){Wagoner}, {Silbergleit}, \&
  {Ortega-Rodr{\'\i}guez}}]{Wagoner2001}
{Wagoner}, R.~V., {Silbergleit}, A.~S., \& {Ortega-Rodr{\'\i}guez}, M. 2001,
  \apjl, 559, L25, \dodoi{10.1086/323655}

\bibitem[{{Williams} {et~al.}(2018){Williams}, {Gliozzi}, \&
  {Rudzinsky}}]{Williams2018}
{Williams}, J.~K., {Gliozzi}, M., \& {Rudzinsky}, R.~V. 2018, \mnras, 480, 96,
  \dodoi{10.1093/mnras/sty1868}

\bibitem[{{Zhang} {et~al.}(2020){Zhang}, {Yan}, \& {Liu}}]{Zhang2020}
{Zhang}, P., {Yan}, J.~Z., \& {Liu}, Q.~Z. 2020, Acta Astronomica Sinica, 61, 2

\bibitem[{{Zhang} {et~al.}(2017){Zhang}, {Zhang}, {Yan}, {Fan}, \&
  {Liu}}]{Zhang2017}
{Zhang}, P., {Zhang}, P.-f., {Yan}, J.-z., {Fan}, Y.-z., \& {Liu}, Q.-z. 2017,
  \apj, 849, 9, \dodoi{10.3847/1538-4357/aa8d6e}

\bibitem[{{Zhou} \& {Wang}(2005)}]{Zhou2005}
{Zhou}, X.-L., \& {Wang}, J.-M. 2005, \apjl, 618, L83, \dodoi{10.1086/427871}

\bibitem[{{Zhou} {et~al.}(2015){Zhou}, {Yuan}, {Pan}, \& {Liu}}]{Zhou2015}
{Zhou}, X.-L., {Yuan}, W., {Pan}, H.-W., \& {Liu}, Z. 2015, \apjl, 798, L5,
  \dodoi{10.1088/2041-8205/798/1/L5}

\bibitem[{{Zoghbi} {et~al.}(2010){Zoghbi}, {Fabian}, {Uttley}, {Miniutti},
  {Gallo}, {Reynolds}, {Miller}, \& {Ponti}}]{Zoghbi2010}
{Zoghbi}, A., {Fabian}, A.~C., {Uttley}, P., {et~al.} 2010, \mnras, 401, 2419,
  \dodoi{10.1111/j.1365-2966.2009.15816.x}

\end{thebibliography}

\end{document}